\documentclass[aps,prb,twocolumn,showpacs]{revtex4}
\usepackage{graphicx}
\usepackage{epsfig}
\usepackage{amsmath}
\usepackage{amssymb}
\usepackage{amsfonts}%
\begin{document} 
\title{Spin polarizations and spin Hall currents in a two-dimensional
electron gas with magnetic impurities}
\author{C. Gorini, P. Schwab, M. Dzierzawa}
\affiliation{Institut f\"ur Physik, Universit\"at Augsburg, 86135 Augsburg, Germany}
\author{R. Raimondi}
\affiliation{CNISM and Dipartimento  di Fisica "E. Amaldi", Universit\`a  Roma Tre, 00146 Roma, Italy}

\begin{abstract}
We consider a two-dimensional electron gas in the presence of Rashba spin-orbit coupling, 
and study the effects of magnetic $s$-wave and long-range non-magnetic
impurities
on the spin-charge dynamics of the system.  We focus on voltage induced 
spin polarizations and their relation to spin Hall currents.
Our results are obtained using the quasiclassical Green function technique, and
hold in the full range of the disorder parameter $\alpha p_F\tau$.
\end{abstract} 

\date{\today} 
\maketitle 

In the field of spintronics, much attention has recently been paid 
to spin-orbit related phenomena in semiconductors.  One such phenomenon
is the spin Hall effect, i.e.~a spin current flowing perpendicular
to an applied electric field \cite{dyakonov1971, hirsh1999, murakami2003, sinova2004}.
It is now well known that for linear-in-momentum spin-orbit
couplings like the Rashba or Dresselhaus ones the spin Hall current vanishes 
exactly in the bulk of a disordered two-dimensional
electron gas (2DEG) \cite{inoue2004, mishchenko2004, raimondi2005, khaetskii2006}.
This can be understood by looking at the peculiar form
of the continuity equations for the spin, as derived
from its equations of motion in operator form \cite{rashba2004, dimitrova2005, chalaev2005}. 
For a \textit{magnetically} disordered 2DEG
things are however different, and a non-vanishing spin Hall conductivity
is found \cite{inoue2006, gorini2008, wang2007}.
Once more, a look at the continuity equations provides a clear and simple
explanation of the effect \cite{gorini2008}: a new term, whose
appearance is due to magnetic impurities, directly relates
in-plane spin polarizations, induced by the electric field, to spin currents.
As the former, which have been the object of both theoretical and
experimental studies \cite{edelstein1990, aronov1989, kato2004, yang2006, sih2005, stern2006}, 
are influenced by the type of non-magnetic scatterers
considered, we forgo the simplified assumption that these be
$s$-wave, and take into account the full angle dependence of the
scattering potential.
Besides going beyond what is currently found in the literature, where, in the presence
of magnetic impurities, the non-magnetic disorder is either neglected or
purely $s$-wave, our approach also shows the interplay 
between polarizations and spin currents in a 2DEG \cite{milletari2008}. 
We note that in the correct limits our results agree with what is found in
Ref.~[\onlinecite{wang2007}].  On the other hand a discrepancy with
Ref.~[\onlinecite{inoue2006}] arises.
    
For the calculations we rely on the Eilenberger equation for
the quasiclassical Green function in the presence of spin-orbit coupling \cite{raimondi2006}.
The spin-orbit energy is taken to be small compared to the Fermi energy, 
i.e. $\alpha p_F\ll\epsilon_F$ - or equivalently $\alpha\ll v_F$ -
and the standard metallic regime condition $1/\tau\ll\epsilon_F$ is also assumed.
Here $\alpha$ is the spin-orbit coupling constant, 
$p_F(v_F)$ the Fermi momentum (velocity) in the absence of such coupling,
and $\tau$ the elastic quasiparticle lifetime due to non-magnetic scatterers.
Our results hold for a wide range of values of the dimensionless parameter
$\alpha p_F\tau$, since this is not restricted by the above assumptions. 
Contributions of order $(\alpha/v_F)^2$ are neglected throughout.
We focus on intrinsic effects in the Rashba model; extrinsic ones
\cite{tse2006}, Dresselhaus terms \cite{trushin2007} and hole gases
\cite{liu2008} are not taken into account.
Finally, weak localization corrections, 
which could in principle play an important role 
\cite{chalaev2005}, are beyond the scope of our present work.

The Hamiltonian of the 2DEG, 
confined to the $x$-$y$ plane, reads
\begin{equation}
\label{hamiltonian}
H = \frac{{\bf p}^2}{2m} - {\bf b}\cdot{\boldsymbol\sigma} + V({\bf x}),
\end{equation}
with ${\bf b}=\alpha{\bf  e}_z\times{\bf p}$ the Rashba internal field,
${\boldsymbol\sigma}$ the vector of Pauli matrices, 
and $V({\bf x})=V_{\mathrm{nm}}({\bf x})+V_{\mathrm{m}}({\bf x})$ the disorder potential
due to randomly distributed impurities
\cite{noteone}.
Non-magnetic scatterers give rise to $V_{\mathrm{nm}}({\bf x})$
\begin{equation}
\label{disorder1}
V_{\mathrm{nm}}({\bf x})=\sum_i\;U({\bf x}-{\bf R}_i),
\end{equation}
while $V_{\mathrm{m}}({\bf x})$ describes magnetic $s$-wave disorder
\begin{equation}
\label{disorder2}
V_{\mathrm{m}}({\bf x})=\sum_i\;{\bf B}\cdot{\boldsymbol\sigma}\delta({\bf x}-{\bf R}_i).
\end{equation}
Both potentials are treated in the Born approximation, and the standard averaging technique is applied.

To begin with, we look at the continuity equation for the $s_y$ spin polarization
\cite{gorini2008, notecontinuity}
\begin{equation}
\label{continuity1}
\partial_ts_y + \partial_{\bf x}\cdot{\bf j}_{s_y}
=
-2m\alpha j^y_{s_z} - \frac{4}{3\tau_{sf}}s_y,
\end{equation}
where the second term on the r.h.s. is due to magnetic impurities.
Here
$\tau_{sf}$ is the spin-flip time which stems from the potential (\ref{disorder2})
[cf. Eq.~(\ref{spinflip})].
Under stationary and uniform conditions the above equation
implies a vanishing spin current - hence a vanishing spin Hall conductivity - 
unless magnetic disorder is also present, in which case instead
\begin{equation}
\label{continuity2}
j^y_{s_z} = - \frac{2}{3m\alpha \tau_{sf}}s_y.
\end{equation} 
Since the out-of-plane polarized spin current is related to the in-plane
spin polarization, we now use simple physical arguments to explain how the latter is generated by an applied voltage
\cite{edelstein1990,culcer2007}.
\begin{figure}
\includegraphics[width=0.45\textwidth]{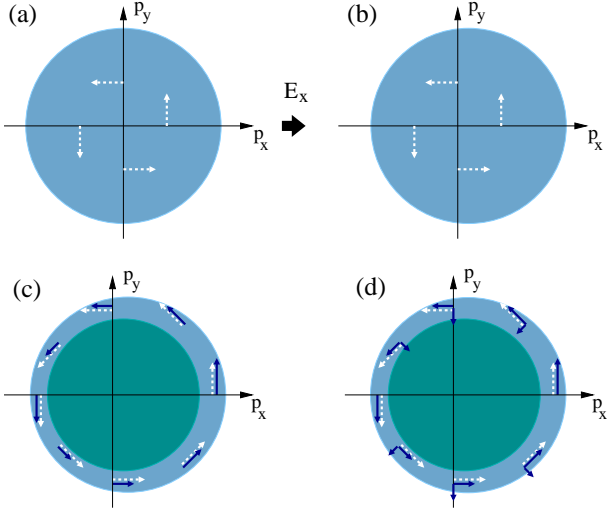}
\caption{(Color online) (a), (b) - The Fermi surface shift, 
$\delta{p}= |e|\mathcal{E}\tau$, due to an applied 
electric field along the $x$-direction.  
The white arrows show the direction of the internal field ${\bf b}$.
(c), (d) -
Shifted bands and spin polarization in stationary conditions.
(c) - Asymmetric shift of the two bands when
angle dependent scattering is present.  The long dark (blue) arrows
show the contributions to the spin polarization arising
from a sector $\mbox{d}\varphi$ of phase space.
(d) - When magnetic disorder is turned on, additional contributions orthogonal to the
internal field ${\bf b}$ appear, here shown by the short inward and
outward pointing (blue) arrows.
Out-of-plane contributions are also present, but for the sake of simplicity
not shown.
}
\label{edelsteinfig}
\end{figure}
Since
the Fermi surface is shifted by an amount proportional to the applied electric
field (say along the $x$-direction), as shown in  Fig.~\ref{edelsteinfig} (a),(b), 
there will be more occupied states with spin up 
- along $y$ - than with spin down.  In the case of short-range disorder, the total in-plane
polarization can be estimated to be proportional
to the density of states times the shift in momentum,
$s_y\sim N\delta p\sim N|e|\mathcal{E}\tau$.
Since in the present situation we are dealing with the two Fermi surfaces corresponding to 
the two helicity bands 
$\epsilon_{\pm}=p^2/2m\pm\alpha p$,  obtained from the Hamiltonian (\ref{hamiltonian}),
one expects $s_y\sim (N_+-N_-)\delta p$, where, for the
Rashba interaction, one has $N_{\pm}=N_0(1\mp\alpha/v_F), N_0=m/2\pi$.
Explicit calculations agree with this simple picture
and lead to the result due to Edelstein 
\cite{edelstein1990},
$s_y=-N_0\alpha|e|\mathcal{E}\tau$.
When long-range disorder is considered, a reasonable guess
could be to substitute for $\tau$ the transport time $\tau_{tr}$
\begin{equation}
\tau\rightarrow\tau_{tr},\;\;\;  \frac{1}{\tau_{tr}}=\int d\theta
W(\theta)(1-\cos  (\theta) ),
\end{equation}
$W(\theta)$ being the angle-dependent scattering probability, so that
$s_y=-N_0\alpha|e|\mathcal{E}\tau_{tr}$.
This was proposed in [\onlinecite{engel2007}], however  
the picture is too simplistic, and therefore 
the guess is wrong.  As discussed in [\onlinecite{milletari2008}], the proper
$s_y$ polarization is given by
$s_y=-N_0\alpha|e|\mathcal{E}\tau_{E}$,
with
\begin{equation}
\label{edelstein time}
\tau\rightarrow\tau_{E},\;\;\;  \frac{1}{\tau_{E}}=\int d\theta
W(\theta)(1-\cos (2\theta) ).
\end{equation}
This particular time $\tau_E$, where $``E"$ stands for
Edelstein, arises from the asymmetric shift of the two Fermi
surfaces, as depicted in Fig.~\ref{edelsteinfig} (c), 
due to different transport times in the two bands.
It shows that contributions from both forward ($\theta=0$)
and backward ($\theta=\pi$) scattering are suppressed.
The next step is to consider what happens when magnetic impurities are included.
Relying once again on the simple picture of the shifted Fermi surface,
one could argue that these have a rather small impact on the spin polarization, 
since the spin-flip scattering time usually makes a small contribution to the total transport time.
However, even when this is the case,
magnetic disorder does not simply modify the total transport time, but has an additional  non-trivial effect.  
In its presence the
spins do not align themselves along the internal ${\bf b}$ field,
since they acquire non-vanishing components in the plane orthogonal
to it - see Fig.~\ref{edelsteinfig} (d).
It is these components who give rise to a finite spin Hall conductivity.
In this respect, magnetic disorder has an effect similar to that
of an in-plane magnetic field: it affects the spin quantization axis
and tilts the spins out of their expected stationary direction.
We now make these arguments quantitative.

The starting point is the Eilenberger equation\cite{raimondi2006},
which we write explicitly for a homogeneous Rashba 2DEG
in linear response to a constant and homogeneous applied electric field
\begin{eqnarray}
  \partial_t g^K
  &=&
  {\bf v}_F\cdot{\bf{\mathcal E}}|e|\partial_{\epsilon}g^K_{eq}
  -\frac{1}{2}\left\{\frac{1}{p_F}\partial_{\varphi}{\bf b}\cdot{\boldsymbol\sigma},
  {\bf e}_{\varphi}\cdot{\bf{\mathcal E}}|e|\partial_{\epsilon}g^K_{eq}\right\}
  \nonumber \\
  &&
  +i\left[{\bf b}\cdot{\boldsymbol\sigma},g^K\right]
  -i\left[\check{\Sigma},\check{g}\right]^K.
  \label{eilenberger1}
\end{eqnarray}
The quasiclassical Green function ($\check{g} \equiv \check{g}_{t_1t_2}(\hat{\bf {p}};{\bf x}) $)
is defined as ($\xi = p^2/2m - \mu$)
\begin{equation}
\label{qc2}
\check{ g } = \frac{i}{\pi} \int {\rm d } \xi \, \check G_{t_1t_2}(\mathbf{ p},\mathbf{ x}), \
\check G = 
\left( \begin{array}{cc} 
G^R & G^K \\
0   & G^A 
\end{array}
\right),
\end{equation} 
where $\check{G}_{t_1t_2}({\bf p},{\bf x})$ is  the Wigner representation of the
Green function, which has a matrix structure in both Keldysh (denoted by the  check symbol) 
and spin space.  
Eq.~(\ref{eilenberger1}) is the equation of motion for the Keldysh component
- the one related to physical observables - identified by
the superscript $``K"$, which will be from now on implicitly assumed and
thus dropped.
Moreover, $g^K_{eq}=\tanh (\epsilon / 2T) (g^R_{eq}-g^A_{eq})$,
where $g^R_{eq}=-g^A_{eq}=1-\partial_{\xi}{\bf
b}\cdot{\boldsymbol\sigma}$, 
indicates the equilibrium - no electric field - function
\cite{raimondi2006}.
All objects are evaluated
at the Fermi surface in the absence of spin-orbit coupling,
while $\varphi$ is the angle defined by the momentum, 
${\bf p}=p(\cos\varphi,\sin\varphi)$, and ${\bf e}_\varphi =(-\sin
\varphi, \cos \varphi)$.
From Eqs.~(\ref{disorder1}) and (\ref{disorder2}) one obtains the 
self-energy contributions 
\begin{equation}
\check{\Sigma}_{\mathrm{nm}}({\bf p})=n_{\mathrm{nm}}\sum_{{\bf p}'}|U({\bf p}-{\bf p}')|^2\check{G}({\bf p}')
\end{equation}
and
\begin{equation}
\check{\Sigma}_{\mathrm{m}}=n_{\mathrm{m}}\frac{B^2}{3}\sum^3_{l=1}\sum_{{\bf p}}\sigma_l\check{G}({\bf p})\sigma_l,
\end{equation}
where $n_{\mathrm{nm}}$ and $n_{\mathrm{m}}$ denote the concentrations 
of non-magnetic and magnetic impurities, respectively. 
In order to consider long-range non-magnetic disorder, we first expand the 
non-magnetic scattering kernel in spherical harmonics of the
scattering angle and neglect
its dependence on the modulus of ${\bf p}$ and ${\bf p}'$
\begin{eqnarray}
n_{\mathrm{nm}}|U|^2 
&=&
\frac{1}{2\pi N_0\tau}\left(1+2K_1\cos(\varphi-\varphi')+ \right. \nonumber \\
&& \left.2K_2\cos(2\varphi-2\varphi')+...\right)
\nonumber \\
&\equiv&
\frac{1}{2\pi N_0\tau}(1+K(\varphi-\varphi'))
\end{eqnarray}
with $\tau$ the non-magnetic contribution to the elastic lifetime.
Then we write the magnetic scattering kernel in terms of  the spin-flip  time $\tau_{sf}$
\begin{equation}
\label{spinflip}
n_{\mathrm{m}} B^2=\frac{1}{2\pi N_0\tau_{sf}}.
\end{equation}
The complete disorder self-energy can then be written separating
its $s$-wave and higher harmonics contributions
\begin{eqnarray}
\check{\Sigma} &=& \check{\Sigma}_{\mathrm{m}} + \check{\Sigma}^1_{\mathrm{nm}} + 
\check{\Sigma}^2_{\mathrm{nm}} \nonumber\\
&=&
-\frac{i}{6\tau_{sf}}\sum^3_{l=1}\sigma_l\langle\check{g}\rangle\sigma_l
-\frac{i}{2\tau}\langle\check{g}\rangle
-\frac{i}{2\tau}\langle K\check{g}\rangle
\end{eqnarray}
where $\langle...\rangle\equiv\int{\rm d}\varphi/2\pi...$.

The connection between $\check{g}$ and the physical observables is made by integrating over
the energy $\epsilon$, which is the Fourier conjugate variable of the time difference
$t_1-t_2$. For instance, the  spin density
is given by the angular average of the Keldysh component \cite{notetwo}
\begin{equation}
\label{qc3b}
{\bf s}={\bf s}^{eq}
-\frac{N_0}{8}\int \mathrm{d}\epsilon \langle\mathrm{Tr}(\boldsymbol{\sigma} g)\rangle.
\end{equation}
In order to solve Eq.~(\ref{eilenberger1}), 
it is convenient to turn it into matrix form,
writing $g$ as a four-vector 
\begin{equation}
g=g_0\sigma_0+{\bf g}\cdot{\boldsymbol\sigma},\;\;
(g_{\mu})=(g_0, {\bf g}).
\end{equation}
Rather than using the standard $(\sigma_x, \sigma_y, \sigma_z)$ basis, we choose to
rotate to $(\sigma_{\parallel}, \sigma_{\perp}, \sigma_z)$, the subscripts
$\parallel$ and $\perp$ indicating respectively the directions
parallel and perpendicular to the internal field ${\bf b}$. 
Defining the rotation matrix ${\bf R}(\varphi)$ by
\begin{eqnarray}
\label{rotation}
\left(
\begin{array}{c}
\sigma_0 \\
\sigma_x \\
\sigma_y \\
\sigma_z
\end{array}
\right)
&=&
\left(
\begin{array}{cccc}
1 & 0 & 0 & 0 \\
0 & \sin\varphi & \cos\varphi & 0 \\
0 & -\cos\varphi & \sin\varphi & 0 \\
0 & 0 & 0 & 1
\end{array}
\right)
\left(
\begin{array}{c}
\sigma_0 \\
\sigma_{\parallel} \\
\sigma_{\perp} \\
\sigma_z
\end{array}
\right),
\end{eqnarray}
one has 
\begin{eqnarray}
&& g_{\mu}' = \sum_{\mu'=0}^{3}{\bf
R}^{-1}_{\mu\mu'}(\varphi)g_{\mu'},  \; \; (g_\mu')= (g_0, g_\parallel, g_\perp, g_z) \\
&& {\bf K}_{\mu\nu}(\varphi,\varphi')
=
\sum_{\mu'=0}^{3}{\bf R}^{-1}_{\mu\mu'}(\varphi)K(\varphi-\varphi'){\bf R}_{\mu'\nu}(\varphi').
\end{eqnarray}
Expanding in harmonics - we also drop the four-vector indices
\begin{eqnarray}
{\bf K}(\varphi,\varphi')=
{\bf K}^{(a)}+\cos(\varphi-\varphi'){\bf K}^{(b)}+\sin(\varphi-\varphi'){\bf K}^{(c)}+... .
\end{eqnarray}
In the above we have defined
\begin{equation}
{\bf K}^{(a)}=
\left(
\begin{array}{cccc}
0 & 0 & 0 & 0 \\
0 & K_1 & 0 & 0 \\
0 & 0 & K_1 & 0 \\
0 & 0 & 0 & 0 
\end{array}
\right),\;
{\bf K}^{(b)}=
\left(
\begin{array}{cccc}
2K_1 & 0 & 0 & 0 \\
0 & K_2 & 0 & 0 \\
0 & 0 & K_2 & 0 \\
0 & 0 & 0 & 2K_1
\end{array}
\right)
\end{equation}
and
\begin{equation}
{\bf K}^{(c)}=
\left(
\begin{array}{cccc}
0 & 0 & 0 & 0 \\
0 & 0 & -K_2 & 0 \\
0 & K_2 & 0 & 0 \\
0 & 0 & 0 & 0
\end{array}
\right).
\end{equation}
For the purpose of calculating polarizations and spin currents 
the higher harmonics play no role and are thus ignored.

By using that $g^R_{eq}=-g^A_{eq}=1-\partial_{\xi}{\bf b}\cdot{\boldsymbol\sigma}$
and performing a rotation to the new spin basis,
one can write Eq.~(\ref{eilenberger1}) as
\begin{equation}
\label{eilenberger2}
\partial_t g' = \frac{1}{\tau^*}\left[-{\bf M}g'+({\bf N}_0+{\bf
N}_1)\langle g'\rangle
+({\bf N}_2+{\bf N}_3)\langle {\bf K}g'\rangle\right] + S_{\mathcal E}.
\end{equation}	
 
The matrices appearing in Eq.~(\ref{eilenberger2}) read
\begin{equation}
{\bf M}
=
\left(
\begin{array}{cccc}
1 & -\frac{\tau^*}{\tau}\frac{\alpha}{v_F}K_1 & 0 & 0 \\
-\frac{\tau^*}{\tau}\frac{\alpha}{v_F}K_1 & 1 & 0 & 0 \\
0 & 0 & 1 & 2\alpha p_F\tau^* \\
0 & 0 & -2\alpha p_F\tau^* & 1
\end{array}
\right)
\end{equation}
\begin{equation}
{\bf N}_0
=
\left(
\begin{array}{cccc}
1 & 0 & 0 & 0 \\
0 & 1-\frac{4\tau^*}{3\tau_{sf}} & 0 & 0 \\
0 & 0 & 1-\frac{4\tau^*}{3\tau_{sf}} & 0 \\
0 & 0 & 0 & 1-\frac{4\tau^*}{3\tau_{sf}}
\end{array}
\right)
\end{equation}
\begin{equation}
{\bf N}_1
=
\frac{\alpha}{v_F}
\left(
\begin{array}{cccc}
0 & -(1-\frac{4\tau^*}{3\tau_{sf}})& 0 & 0 \\
-1 & 0 & 0 & 0 \\
0 & 0 & 0 & 0 \\
0 & 0 & 0 & 0
\end{array}
\right)
\end{equation}
\begin{equation}
{\bf N}_2
=
\frac{\tau^*}{\tau}
\frac{\alpha}{v_F}
\left(
\begin{array}{cccc}
0 & -1 & 0 & 0 \\
-1 & 0 & 0 & 0 \\
0 & 0 & 0 & 0 \\
0 & 0 & 0 & 0
\end{array}
\right),\;
{\bf N}_3
=
\frac{\tau^*}{\tau}
\left(
\begin{array}{cccc}
1 & 0 & 0 & 0 \\
0 & 1 & 0 & 0 \\
0 & 0 & 1 & 0 \\
0 & 0 & 0 & 1
\end{array}
\right)
\end{equation}
where $\tau^*$ is the elastic quasi-particle life time, defined as
\begin{equation}
\frac{1}{\tau^*}\equiv\frac{1}{\tau}+\frac{1}{\tau_{sf}}
\end{equation}
which we now use for convenience of notation, but will be later incorporated into
the proper transport time.
Finally, $S_{\mathcal E}$ is the source term due to the electric field.
We take this to be along the $x$-direction, so that
\begin{equation}
S_{\mathcal{E}}\equiv
|e|v_F\mathcal{E}\partial_{\epsilon}(2\tanh(\epsilon/2T))
\left(
\begin{array}{c}
\cos\varphi \\
-\cos\varphi \frac{\alpha}{v_F} \\
-\sin\varphi \frac{\alpha}{v_F} \\
0
\end{array}
\right).
\end{equation}

\begin{figure}
\flushleft{(a)}
\\
\includegraphics[width=0.44\textwidth]{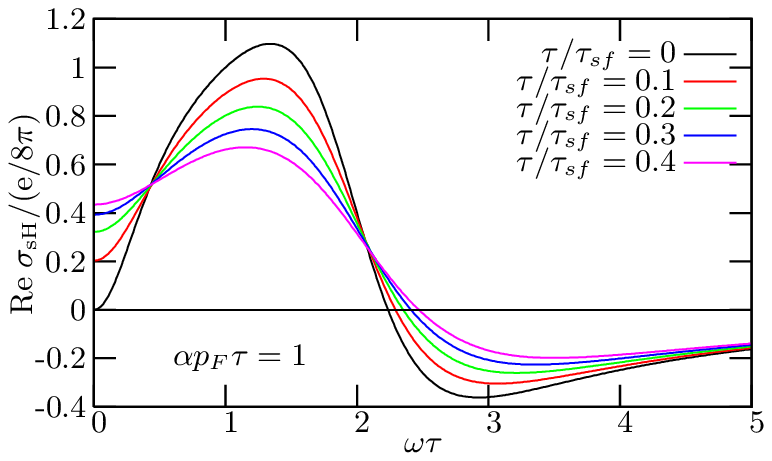}
\flushleft{(b)}
\\
\includegraphics[width=0.45\textwidth]{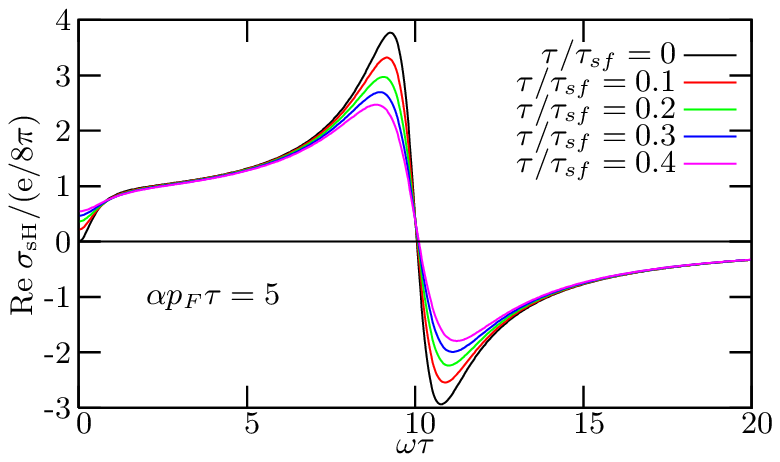}
\caption{(Color online) Real part of the frequency dependent spin Hall 
conductivity in units of the universal value $|e|/8\pi$
for $\alpha p_F\tau=1$ (a) and $\alpha p_F\tau=5$ (b).
The different curves correspond to different values of the ratio
$\tau/\tau_{sf} = 0, 0.1, 0.2, 0.3, 0.4$ 
(from top to bottom at the maximum of Re~$\sigma_{sH}$).}
\label{spinhall}
\end{figure}

Solving for the $s_z$ spin current flowing along $y$,  
we obtain
\begin{eqnarray}
j^y_{s_z}&=& -\frac{N_0}{4}\int\,\mbox{d}\epsilon\, v_F\langle\hat{\bf p}_yg_z\rangle \nonumber \\
&=&
-\frac{N_0}{4}\int\,\mbox{d}\epsilon\,
\left[-\frac{\frac{4}{3\tau_{sf}}-i\omega}{2m\alpha}\right]
(\langle\hat{\bf p}_yg_{\perp}\rangle-\langle\hat{\bf p}_xg_{\parallel}\rangle)
\nonumber \\
&=&
-\frac{N_0}{4}\int\,\mbox{d}\epsilon\,
\left[-\frac{\frac{4}{3\tau_{sf}}-i\omega}{2m\alpha}\right]\langle g_y\rangle
\nonumber \\
&=& 
\label{continuity3}
\left[-\frac{\frac{4}{3\tau_{sf}}-i\omega}{2m\alpha}\right]s_y,	
\end{eqnarray}
i.e.~the continuity equation result, Eq.~(\ref{continuity1}), under homogeneous conditions.
In the third line we have used Eq.~(\ref{rotation}) to set
$\langle g_y\rangle=\langle\hat{\bf p}_yg_{\perp}\rangle-\langle\hat{\bf p}_xg_{\parallel}\rangle$.
Similarly, one obtains the complete expression for the frequency dependent $s_y$ spin polarization
\begin{eqnarray}
s_y &=& -N_0\alpha|e|\mathcal{E}\;2(\alpha p_F)^2 \nonumber \\
&&
\times \left[
\left(\frac{1}{\tau_{tr}}-i\omega\right)\left(\frac{1}{\tau_E}-i\omega\right)
\left(\frac{4}{3\tau_{sf}}-i\omega\right)
+ \right. \nonumber \\
&&
\label{frequency edelstein}
\left. +\, 2(\alpha p_F)^2\left(\frac{1}{\tau_E}+\frac{4}{3\tau_{sf}}-2i\omega\right)
\right]^{-1}.
\end{eqnarray}
Besides $1/\tau_{sf}$, 
there appear in the above two other different time scales
\begin{eqnarray}
&&
\frac{1}{\tau_{tr}}\equiv\frac{1}{\tau}(1-K_1)+\frac{1}{\tau_{sf}},\;\;\;\;
\frac{1}{\tau_E}\equiv\frac{1}{\tau}(1-K_2)+\frac{1}{\tau_{sf}}.\;\; \nonumber
\end{eqnarray}
The first, $\tau_{tr}$, is the total transport time.  
The second, $\tau_E$, is the generalization of the characteristic time 
related to the $s_y$ spin polarization introduced in (\ref{edelstein time}).
By using Eq.~(\ref{frequency edelstein}) in Eq.~(\ref{continuity3}), one
obtains the expression for the frequency dependent spin Hall conductivity
\begin{eqnarray}
\sigma_{sH}(\omega) &=&
\frac{|e|}{4\pi}
\left(\frac{4}{3\tau_{sf}}-i\omega\right)\;2(\alpha p_F)^2 \nonumber\\
&&
\times \left[
\left(\frac{1}{\tau_{tr}}-i\omega\right)\left(\frac{1}{\tau_E}-i\omega\right)
\left(\frac{4}{3\tau_{sf}}-i\omega\right)
+ \right. \nonumber \\
&&
\left. + \,2(\alpha p_F)^2\left(\frac{1}{\tau_E}+\frac{4}{3\tau_{sf}}-2i\omega\right)
\right]^{-1}.
\end{eqnarray}
Its real part is displayed in Fig.~\ref{spinhall} for different
values of the disorder parameter $\alpha p_F\tau$.
In the limit $\omega\rightarrow0$, the magnitude of the spin Hall
conductivity depends on the value of $\alpha p_F\tau$ as well as
on the ratio $\tau/\tau_{sf}$.  
In the absence of magnetic impurities one has the 
known result $\sigma_{sH}=0$. 
As spin flip scattering grows, the
conductivity reaches values of the order of the ``universal" $|e|/8\pi$.
This was noted already in [\onlinecite{inoue2006}], where however, as pointed out in the beginning,
angle dependent scattering was not considered. 
Large values of $\alpha p_F\tau$ can be achieved
both in III-V and II-VI semiconducting materials.
Doping the latter with Mn allows to control the spin-flip 
time $\tau_{sf}$ while only weakly affecting the electrons
mobility \cite{gui2004, daumer2003, crooker1995},
even though it is not perfectly clear whether these
can appropriately be described in terms of the linear Rashba model
\cite{noteHgTe}.
Additionally, for certain frequencies one can see crossing points
[$\omega \tau \approx 0.5 $ and $\omega \tau \approx 2$ in
Fig.\ \ref{spinhall} (a)] at which magnetic disorder has no effect
on the spin Hall response.  Such points are well defined only
when $\alpha p_F\tau\approx1$.
For clean ($\alpha p_F\tau\gg1$) 
or dirty ($\alpha p_F\tau\ll1$) samples 
the different curves cross each other
over a progressively wider range of frequencies.

Finally, in the diffusive regime, $\omega\tau_{tr}\ll1, \alpha p_F\tau_{tr}\ll1$, and
assuming $\tau_{tr}/\tau_{sf}\ll1, \tau_{E}/\tau_{sf}\ll1$, one obtains the
following spin-diffusion equations
\begin{eqnarray}
\partial_t s_x &=& -\left(\frac{1}{\tau_s}+\frac{4}{3\tau_{sf}}\right)s_x
\\ \label{eq35}
\partial_t s_y &=& -\left(\frac{1}{\tau_s}+\frac{4}{3\tau_{sf}}\right)s_y
-\alpha N_0|e|\mathcal{E}\frac{\tau_E}{\tau_s}
\\
\partial_t s_z &=& -\left(\frac{2}{\tau_s}+\frac{4}{3\tau_{sf}}\right)s_z
\end{eqnarray}
where
$(2\alpha p_F\tau_{tr})^2/2\tau_{tr}\equiv1/\tau_s$
is the Dyakonov-Perel spin relaxation rate, tied to Rashba spin-orbit
coupling.
From Eq.~(\ref{eq35}) the sensitivity of the in-plane spin
polarization on spin-flip scattering is apparent: in the stationary
limit the source (proportional to $\mathcal E$) is balanced by the spin relaxation.
Spin-flip scattering leaves the source unchanged, whereas it enhances
the relaxation rate so that in the end $s_y$ is reduced.

In conclusion, we studied the combined effect of long-range
and magnetic disorder on voltage induced spin polarizations
and the related spin Hall currents in a Rashba 2DEG.
We investigated homogeneous but non-static conditions,
from the dirty ($\alpha p_F\tau\ll1$)
to the clean ($\alpha p_F\tau\gg1$) regime. 
Care is required when treating long-range disorder 
because of the two-band structure of the problem, 
while magnetic impurities, even in low concentrations, 
play a non-trivial role beyond that of a simple redefinition of the time scales.

This work was supported by the Deutsche Forschungsgemeinschaft through
SFB 484 and SPP 1285 and by CNISM under Progetti Innesco 2006.


\begin{thebibliography}{999}
\bibitem{notetwo}Within the quasiclassical formalism, $g$ describes the dynamical part only, to which
the equilibrium part has to be added.
\bibitem{noteHgTe} More precisely, the Rashba Hamiltonian is appropriate for narrow quantum wells 
(width $\lesssim 6 {\mbox nm}$), but most likely not for wider structures, in which the so-called
inverted-band structure manifests itself.
\end{thebibliography}
\end{document}